\documentclass[10pt, conference]{IEEEtran}
\IEEEoverridecommandlockouts
\usepackage{balance}
\usepackage{cite}
\usepackage{fancyhdr}
\usepackage{amssymb}
\usepackage{amsmath}
\usepackage{multirow}
\usepackage{graphicx}
\usepackage{algorithm}
\usepackage{algorithmic}
\usepackage{url}
\usepackage{booktabs}
\usepackage{color}
\usepackage{moreverb}
\usepackage{balance}
\usepackage{amssymb,amsmath}
\usepackage{wrapfig}
\usepackage{multirow}
\usepackage{graphicx}
\usepackage{algorithm}
\usepackage{algorithmic}
\usepackage{epstopdf}
\usepackage{subfigure}
\usepackage{subfig}
\usepackage{url}
\usepackage{enumitem}
\usepackage{miniplot}
\usepackage{ifthen}

\usepackage{pifont}
\usepackage{bbding}
\usepackage{diagbox}
\usepackage{ragged2e}
\usepackage{tikz}
\usepackage{etoolbox}
\usepackage{subfigure}
\usepackage{textcomp}
\newcommand*{\circled}[1]{\lower.7ex\hbox{\tikz\draw (0pt, 0pt)%
    circle (.5em) node {\makebox[1em][c]{\small #1}};}}
\robustify{\circled}

\renewcommand{\raggedright}{\leftskip=0pt \rightskip=0pt plus 0cm}
\newboolean{showcomments}
\setboolean{showcomments}{true}
\ifthenelse{\boolean{showcomments}}
{ \newcommand{\mynote}[2]{
    \fbox{\bfseries\sffamily\scriptsize#1}
    {\small$\blacktriangleright$\textsf{\emph{#2}}$\blacktriangleleft$}}}
{ \newcommand{\mynote}[2]{}}

\ifthenelse{\boolean{showcomments}}
{
  
}{
  \newcommand{\mynote}[2]{}
}

\definecolor{orange}{rgb}{1,0.5,0}

\title{A Differential Testing Approach for Evaluating Abstract Syntax Tree Mapping Algorithms}

\author{\IEEEauthorblockN{Yuanrui Fan\IEEEauthorrefmark{1}\IEEEauthorrefmark{2}, Xin Xia\IEEEauthorrefmark{3}\IEEEauthorrefmark{4}\thanks{\IEEEauthorrefmark{4}Corresponding author.}, David Lo\IEEEauthorrefmark{5}, Ahmed E. Hassan\IEEEauthorrefmark{6}, Yuan Wang\IEEEauthorrefmark{7}, Shanping Li\IEEEauthorrefmark{1}}
\IEEEauthorblockA{\IEEEauthorrefmark{1}Zhejiang University, China; \IEEEauthorrefmark{2}PengCheng Laboratory, China; \IEEEauthorrefmark{3}Monash University, Australia; \\
\IEEEauthorrefmark{5}Singapore Management University, Singapore; \IEEEauthorrefmark{6}Queen's University, Canada; \IEEEauthorrefmark{7}Huawei Sweden Research Center \\
\ \{yrfan, shan\}@zju.edu.cn, Xin.Xia@monash.edu, davidlo@smu.edu.sg, ahmed@cs.queensu.ca, Yuan.Wang1@huawei.com}}

\begin{document}
\maketitle

\begin{abstract}
Abstract syntax tree (AST) mapping algorithms are widely used to analyze changes in source code.
Despite the foundational role of AST mapping algorithms, little effort has been made to evaluate the accuracy of AST mapping algorithms, i.e., the extent to which an algorithm captures the evolution of code.
We observe that a program element often has only one best-mapped program element.
Based on this observation, we propose a hierarchical approach to automatically compare the similarity of mapped statements and tokens by different algorithms.
By performing the comparison, we determine if each of the compared algorithms generates inaccurate mappings for a statement or its tokens.
We invite 12 external experts to determine if three commonly used AST mapping algorithms generate accurate mappings for a statement and its tokens for 200 statements.
Based on the experts' feedback, we observe that our approach achieves a precision of 0.98--1.00 and a recall of 0.65--0.75.
Furthermore, we conduct a large-scale study with a dataset of ten Java projects containing a total of 263,165 file revisions.
Our approach determines that GumTree, MTDiff and IJM generate inaccurate mappings for 20\%--29\%, 25\%--36\% and 21\%--30\% of the file revisions, respectively.
Our experimental results show that state-of-the-art AST mapping algorithms still need improvements.
\end{abstract}

\begin{IEEEkeywords}
Program element mapping, abstract syntax trees, software evolution
\end{IEEEkeywords}

\section{Introduction}\label{intro}


Program element mapping algorithms are the underlying basis for analyzing changes between two versions of a source code file (i.e., a file revision)~\cite{kim2006program}.
Abstract syntax tree (AST) mapping algorithms represent a file revision and program elements of the file as two abstract syntax trees (ASTs) and nodes, respectively~\cite{falleri2014fine, dotzler2016move, frick2018generating}.
The algorithms approximate the similarity of nodes and calculate mappings of nodes between the two ASTs.
We define accurate mappings as mappings that can reflect the evolution of code well.


Edit actions including adding, deleting, moving and updating nodes of an AST can be calculated based on the generated mappings by an AST mapping algorithm~\cite{chawathe1996change}.
Such edit actions can describe changes to the syntactic structure of code, e.g., parameters added in a method call can be represented as added nodes in an AST.
Accurate mappings lead to accurate edit actions that can reflect a developer's intent.
Many prior studies apply AST mapping algorithms to calculate edit actions for further analyses, e.g., API recommendation~\cite{nguyen2016api}, mining code change patterns~\cite{li2018logtracker}, and automated program repair~\cite{tufano2019empirical}.
The accuracy of the used AST mapping algorithms is vital for the correctness of the proposed approaches by prior studies.



However, evaluations of the accuracy of AST mapping algorithms are limited.
When evaluating the generated mappings by an algorithm, prior work relies heavily on manual analysis of the derived edit actions from the mappings~\cite{falleri2014fine, dotzler2016move, frick2018generating}.
Since it is infeasible to analyze all file revisions by hand, prior studies select a small sample for analysis.
The many cases where AST mapping algorithms perform badly cannot be revealed.
Furthermore, manually analyzing the mappings of program elements is time-consuming and tedious.
An approach that can automatically find the inaccurate mappings as generated by AST mapping algorithms would be helpful.
Practitioners and researchers can leverage such an approach to explore and navigate the generated mappings by an algorithm before performing further analyses.

In this paper, we propose an approach for evaluating AST mapping algorithms.
We observe that a program element $e$ often has only one best-mapped program element $\hat{e}$ in a file revision.
The element $\hat{e}$ might be empty, i.e., $e$ should not be mapped.
If two algorithms inconsistently map such an element, at least one of the algorithms inaccurately mapped the element.
We define \emph{similarity} of two program elements in a file revision as the likelihood of the two elements to be mapped.
Our idea is that if an algorithm maps a more similar program element for $e$ than another algorithm, the latter algorithm is likely to have made a mistake.
We aim to automatically compare the similarity of mapped program elements by different AST mapping algorithms.
In the software testing area, this approach is referred to as \emph{differential testing}~\cite{mckeeman1998differential}.




To ease analysis, we refine the mappings of AST nodes into mappings of each statement and tokens in the statement.
And we treat each statement as an analysis unit.
By doing so, we avoid analyzing the mappings of AST nodes at all granularity levels.
Notice that statements include declarations (e.g., method declarations) in our paper.
A token is defined as a sequence of characters representing a program element such that none of its subsequences represents a program element.
Given two algorithms, if they inconsistently map a statement or its tokens, we refer to such statements as \emph{statements with inconsistent mappings} for the two algorithms.
If an algorithm inaccurately maps a statement or its tokens, we refer to such statements as \emph{statements with inaccurate mappings} for the algorithm.


We analyzed three AST mapping algorithms in our paper, namely GumTree~\cite{falleri2014fine}, MTDiff~\cite{dotzler2016move} and IJM (Iterative Java Matcher)~\cite{frick2018generating}.
As shorthand notations, we use GT and MTD to represent GumTree and MTDiff, respectively.
We manually analyze 575 statements with inconsistent mappings for comparing the algorithms.
Through the manual analysis, we design a hierarchical approach to automatically compare the similarity of mapped statements and tokens by different algorithms.
This hierarchical approach uses six measures collectively to perform the comparison.
By performing the comparison, we determine statements with inaccurate mappings for each of the compared algorithms.

We invite 12 external experts to determine if the studied algorithms generate inaccurate mappings (at the statement or token level) for 200 statements.
Compared to the experts' evaluation, we find that our approach achieves a precision of 0.98--1.00 and a recall of 0.65--0.75.
We run the studied algorithms on all the file revisions of ten Java projects.
We use our approach to analyze the generated mappings by the algorithms.
For 20\%--29\%, 25\%--36\% and 21\%--30\% of the file revisions, GT, MTD and IJM are determined to generate inaccurate mappings, respectively.
The results show that state-of-the-art AST mapping algorithms still need improvements.
We make our code and data publicly available on~\cite{impl}.

Our contributions are summarized as follows:

\begin{itemize}
\item We propose an approach that can automatically detect statements with inaccurate mappings for AST mapping algorithms.
      Almost all of the statements with inaccurate mappings as determined by our approach are also determined as such by experts (98\%--100\%).

\item We use our approach to analyze the generated mappings by GT, MTD and IJM for 263,165 file revisions.
      The three algorithms are determined to generate inaccurate mappings for a considerable number of file revisions.
\end{itemize}





\section{Preliminaries}\label{preliminary}

\noindent\textbf{Abstract syntax tree.}
A source code file can be parsed as an abstract syntax tree (AST).
An AST is a labeled ordered rooted tree, which is composed of a set of nodes that are connected by edges.
An edge represents a parent-child relationship.
A node $n_1$ is the \emph{parent} of another node $n_2$, if $n_2$ is a child of $n_1$.
The node that has no parent is called the \emph{root node}.
A node that has no child is called a \emph{leaf node}.
For a node, the nodes along the path to the root node are called its \emph{ancestors}.
And the node is called their \emph{descendant}.
Each node in an AST represents a code element (e.g., a statement) with a \emph{label} to indicate its type.
Some nodes have a \emph{value} to indicate the corresponding tokens of the element.

\begin{figure}
  \centering
  \includegraphics[width=0.48\textwidth]{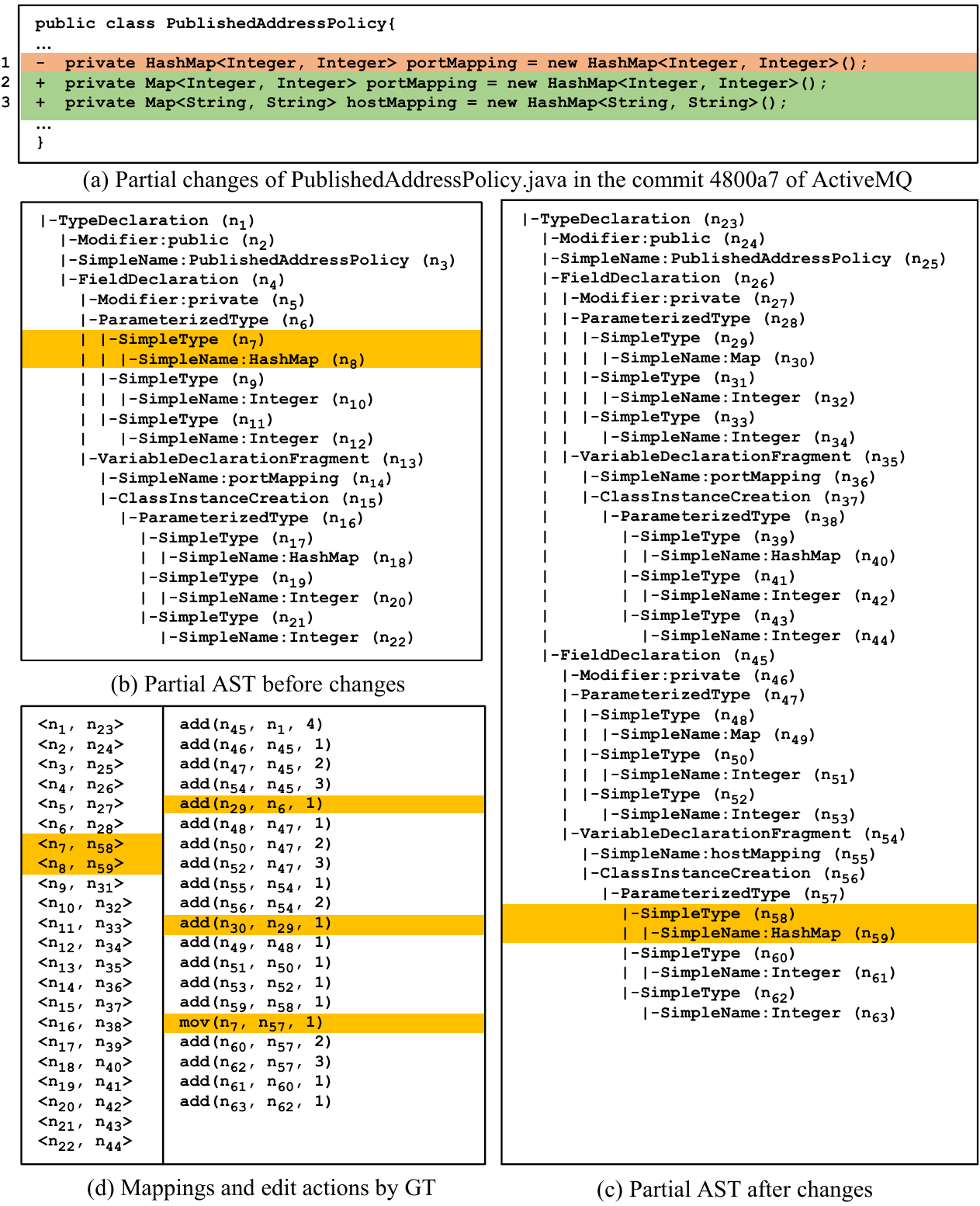}\\
  \caption{Mappings of AST nodes and edit actions as calculated by GT for the partial changes of a file in the commit 4800a7 of ActiveMQ.}\label{fig:mapping-example}
\end{figure}

\underline{\emph{Example 2.1.}} Fig.~\ref{fig:mapping-example}(b) and Fig.~\ref{fig:mapping-example} present the two ASTs before and after the code changes at lines 1--3 in Fig.~\ref{fig:mapping-example}(a).
The ASTs are built using GT~\cite{falleri2014fine}.
We only show partial ASTs for clarity.
The AST in Fig.~\ref{fig:mapping-example}(b) contains 22 nodes.
The node $n_{4}$ has three child nodes $n_5$, $n_6$ and $n_{13}$, and its label is \texttt{FieldDeclaration}.
The label of $n_5$ is \texttt{Modifier}, and it has a string value \texttt{private} indicating the field declaration is private for the class.

\noindent\textbf{AST mapping algorithms.}
Given a file revision, an AST mapping algorithm parses the file before and after the revision as two ASTs.
Let us denote the two ASTs as \emph{source AST} and \emph{target AST}, respectively.
By approximating the similarity of nodes, the algorithm calculates mappings of nodes between the two ASTs.
Only the nodes that have the same label can be mapped.
The mappings are a set of pairs $\langle n_s, n_t \rangle$, in which $n_s$ belongs to the source AST and $n_t$ belongs to the target AST.

\underline{\emph{Example 2.2.}} We leverage GT to calculate mappings of nodes between the ASTs shown in Fig.~\ref{fig:mapping-example}(b) and Fig.~\ref{fig:mapping-example}(c).
Fig.~\ref{fig:mapping-example}(d) presents the mappings in the first column.
Specially, $n_4$ is mapped to $n_{26}$, indicating that the field declaration \texttt{portMapping} at line 1 in Fig.~\ref{fig:mapping-example}(a) is mapped to the field declaration \texttt{portMapping} at line 2 in Fig.~\ref{fig:mapping-example}(a).
In addition, $n_7$ and $n_8$ are mapped to $n_{58}$ and $n_{59}$, respectively.
Such mappings indicate that the first \texttt{HashMap} of line 1 and the \texttt{HashMap} of line 3 in Fig.~\ref{fig:mapping-example}(a) are mapped, i.e., GT considers that the two \texttt{HashMap} tokens are the same element across the file revision.

\noindent\textbf{Edit actions.}
Based on mappings of AST nodes, a series of edit actions can be calculated to transform the source AST to the target AST~\cite{chawathe1996change}.
Prior studies apply the Chawathe et al.' algorithm~\cite{chawathe1996change} to calculate the edit actions~\cite{falleri2014fine, dotzler2016move, frick2018generating}.
Generally, there are four types of edit actions:

\begin{itemize}
\item $upd(n, v)$ replaces the value of the node $n$ with a value $v$.

\item $add(n, p, i)$ adds a node $n$ as the $i^{th}$ child of the node $p$ if $p$ is not null.
      Otherwise, the node $n$ is added as the new root node.

\item $del(n)$ deletes a leaf node $n$.

\item $mov(n, p, i)$ moves the node $n$ as the $i^{th}$ child of the node $p$.
      The subtree rooted at $n$ is moved together with $n$.

\end{itemize}


\underline{\emph{Example 2.3.}} Fig.~\ref{fig:mapping-example}(d) presents the edit actions that are calculated based on the mappings shown in the same figure.
We use GT's implementation of Chawathe et al.'s algorithm.
GT produces a sequence of 20 edit actions.
One of the 20 actions is $mov(n_7, n_{57}, 1)$.
This action moves the first \texttt{HashMap} of line 1 in Fig.~\ref{fig:mapping-example}(a) to line 3 in Fig.~\ref{fig:mapping-example}(a).


\noindent\textbf{Current evaluations of AST mapping algorithms.}
Current evaluations of AST mapping algorithms include automatic and manual evaluations~\cite{falleri2014fine, dotzler2016move, frick2018generating}.
They are both based on the edit actions that are calculated from the generated mappings by different algorithms.

The number of edit actions is commonly used as an automatic measure for estimating the cognitive load for a developer when understanding the essence of a file revision~\cite{falleri2014fine, dotzler2016move, frick2018generating}.
Mappings with fewer edit actions are considered to be better.
However, the number of edit actions cannot reflect the accuracy of the mappings~\cite{frick2018generating}.
Prior work relies heavily on manual analyses of the edit actions to determine the accuracy of the mappings~\cite{falleri2014fine, dotzler2016move, frick2018generating}.
Frick et al.~\cite{frick2018generating} proposed three criteria for determining if a mapping is accurate:
(1) each mapping should be comprehensible, i.e., why the two program elements should be mapped;
(2) the generated edit actions should be helpful in understanding the changes;
(3) there exists no other comprehensible mappings resulting in fewer actions.

\underline{\emph{Example 2.4.}}
In Fig.~\ref{fig:mapping-example}(d), we determine the accuracy of the mappings and highlight the inaccurate mappings with a yellow background.
The mappings $\langle n_7, n_{58} \rangle$ and $\langle n_8, n_{59}\rangle$ are determined to be inaccurate.
The nodes in the ASTs and edit actions that relate to the inaccurate mappings are also highlighted.
We consider that it is not comprehensible to map the first \texttt{HashMap} of line 1 to the \texttt{HashMap} of line 3 in Fig.~\ref{fig:mapping-example}(a).
Also, the action $mov(n_7, n_{57}, 1)$ is not helpful in understanding the changes.
It is better to map $n_7$ and $n_8$ to $n_{29}$ and $n_{30}$, i.e., map the first \texttt{HashMap} of line 1 and the \texttt{Map} of line 2 in Fig.~\ref{fig:mapping-example}(a).
Thus, the mappings for the nodes $n_{29}$ and $n_{30}$ are also considered to be inaccurate.

\section{Motivation}\label{motivation}

%
%

Prior evaluations of AST mapping algorithms are limited to manually analyzing a small sample of file revisions~\cite{falleri2014fine, dotzler2016move, frick2018generating}.
For instance, to evaluate GT, Falleri et al. manually analyzed 144 file revisions~\cite{falleri2014fine}.
To evaluate MTD, Dotzler et al. manually analyzed 10 file revisions~\cite{dotzler2016move}.
The many cases where an AST mapping algorithm performs badly cannot be revealed.
An approach that can automatically find inaccurate mappings as generated by an algorithm would be helpful for both researchers and practitioners.

We observe that for a file revision, a program element $e_1$ often has only one best-mapped element $e_2$.
The element $e_2$ can be empty, indicating that $e_1$ should not be mapped.
For instance, in Fig.~\ref{fig:mapping-example}(a), the first \texttt{HashMap} of line 1 should be mapped to the \texttt{Map} of line 2.
If two algorithms inconsistently map such a program element, at least one of the algorithms inaccurately maps the element.
The inaccurate mappings can be found by analyzing the inconsistencies between the two algorithms.

\begin{figure}
  \centering
  \includegraphics[width=0.48\textwidth]{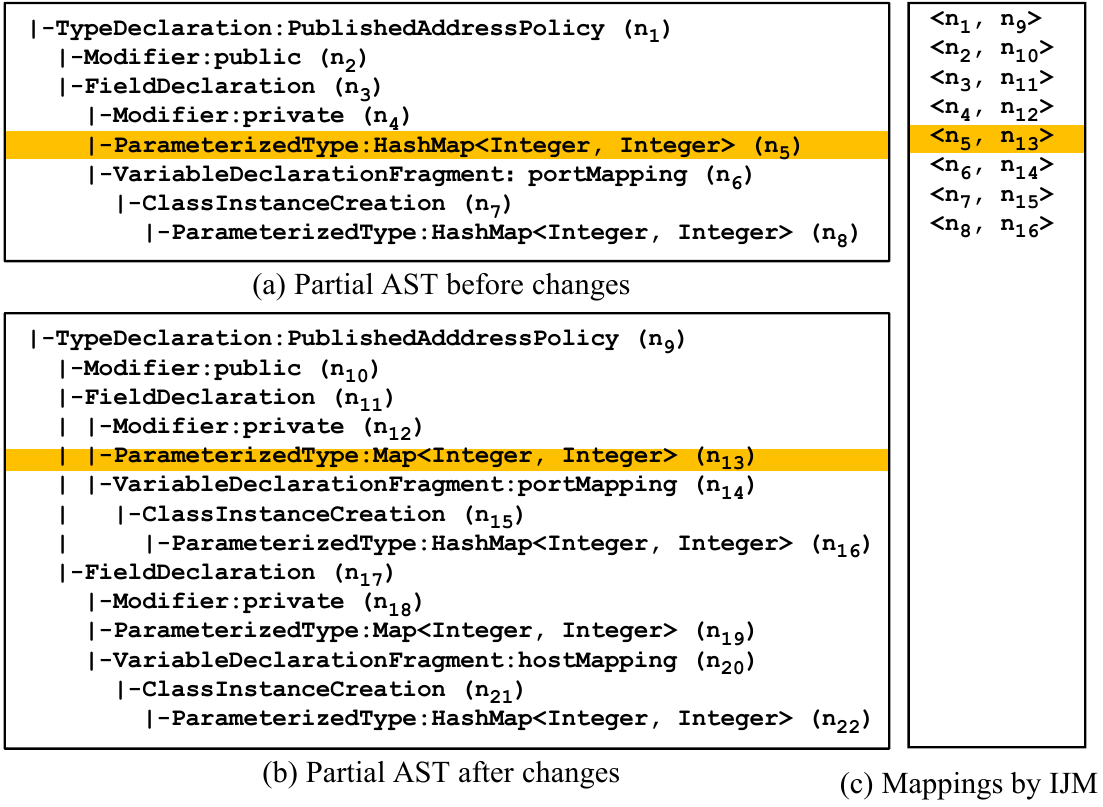}\\
  \caption{Mappings of AST nodes calculated by IJM for the changes shown in Fig.~\ref{fig:mapping-example}(a).}\label{fig:mappings-ijm}
\end{figure}

\begin{figure}
  \centering
  \includegraphics[width=0.48\textwidth]{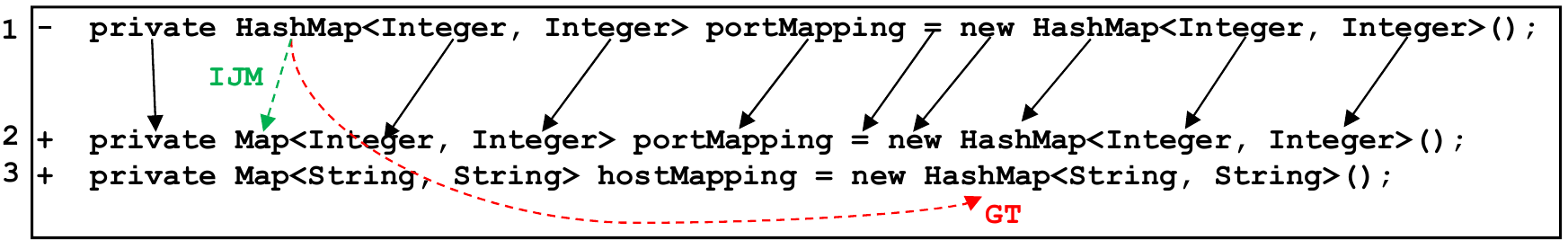}\\
  \caption{The generated mappings by GT and IJM in Fig.~\ref{fig:mapping-example} and Fig.~\ref{fig:mappings-ijm}. Dashed lines show the inconsistent mappings.}\label{fig:token-mappings}
\end{figure}

We provide an example.
For the file revision shown in Fig.~\ref{fig:mapping-example}(a), we use IJM to calculate mappings of AST nodes.
Different AST mapping algorithms may use different ASTs to represent the same file.
Fig.~\ref{fig:mappings-ijm}(a) and (b) show the IJM's ASTs that represent the changes at lines 1-3 in Fig.~\ref{fig:mapping-example}(a).
Fig.~\ref{fig:mappings-ijm}(c) presents the generated mappings by IJM.
The value of $n_5$ denotes the first \texttt{HashMap<Integer, Integer>} of line 1 in Fig.~\ref{fig:mapping-example}(a).
The value of $n_{13}$ denotes the \texttt{Map<Integer, Integer>} of line 2 in Fig.~\ref{fig:mapping-example}(a).
As shown in Fig.~\ref{fig:mappings-ijm}(c), IJM maps $n_5$ to $n_{13}$.
By further mapping the tokens in the value of the two nodes, we find that IJM accurately maps the first \texttt{HashMap} of line 1 in Fig.~\ref{fig:mapping-example}(a).
This mapping is inconsistent with the generated mappings by GT.
Fig.~\ref{fig:token-mappings} visualizes the generated mappings by GT and IJM and their inconsistent mappings.

We find that both GT and IJM map the statement at line 1 to the statement at line 2 in Fig.~\ref{fig:token-mappings}.
The first \texttt{HashMap} of line 1 and the \texttt{Map} of line 2 belong to the mapped statements.
They both denote the type of the field \texttt{portMapping}.
On the other hand, the first \texttt{HashMap} of line 1 and the \texttt{HashMap} of line 3 belong to unmapped statements.
Thus, in comparison to the \texttt{HashMap} of line 3, the \texttt{Map} of line 2 is more similar to the first \texttt{HashMap} of line 1.
Finally, the inaccurately mapped \texttt{HashMap} by GT is detected.
In this motivational example, we attempt to \emph{automatically} perform the above analysis and find the inaccurate mappings as generated by AST mapping algorithms.

\section{Studied AST Mapping Algorithms}\label{algorithm}

In this study, we analyze three state-of-the-art AST mapping algorithms, namely GumTree~\cite{falleri2014fine}, MTDiff~\cite{dotzler2016move} and IJM~\cite{frick2018generating}.
We briefly introduce the three algorithms below.

\textbf{GumTree (GT)} is proposed by Falleri et al~\cite{falleri2014fine}.
Given two ASTs, GT matches nodes between the ASTs in two phases:
In the first phase, GT applies a greedy top-down algorithm to search and map identical subtrees.
In the second phase, GT applies a bottom-up algorithm to map a pair of nodes between the two ASTs if they share a significant number of mapped descendants.
Then, GT tries to map previously unmapped descendants of those nodes.

\textbf{MTDiff (MTD)} is proposed by Dotzler et al.~\cite{dotzler2016move}.
MTD is based on the ChangeDistiller algorithm~\cite{fluri2007change}.
First, MTD applies the identical subtree optimization to reduce the mapping problem by removing unchanged subtrees from the ASTs.
Then, MTD maps nodes using the ChangeDistiller algorithm.
Another four optimizations are finally applied to find additional mappings of nodes.

\textbf{IJM} is proposed by Frick et al.~\cite{frick2018generating}.
IJM is an AST mapping algorithm specialized for Java.
In comparison to GT, IJM works on a reduced AST, in which many name nodes are pruned and the value of each pruned node is merged to its parent node.
Then, IJM splits the AST into parts along each declaration.
Finally, it maps AST nodes from the corresponding parts between the two ASTs using an adaptation of the GT algorithm.
The adaptation adds name-awareness to GT, which considers similarity of values when mapping two nodes.

In this study, we apply the implementations of the three algorithms that are provided on GitHub~\cite{gumtreeGitHub, mtdiffGitHub, ijmGitHub}.

\section{Approach}\label{method}

In this section, we describe the details of our approach.
Notice that our approach is general.
We implement it for Java programs but we can modify it to support other programs as long as we can generate ASTs from them.

The granularity of AST nodes ranges from a statement to a single literal.
It is too complex to analyze mappings of AST nodes of all granularity levels.
Furthermore, finding the corresponding nodes from the used ASTs by different algorithms is a big challenge.
Because different algorithms may use ASTs with different sets of nodes (including leaf nodes) to represent the same file, as shown in Fig.~\ref{fig:mapping-example} and Fig.~\ref{fig:mappings-ijm}.
We observe that different algorithms commonly encode statements as AST nodes.
Additionally, tokens of a file are not impacted by the composing nodes of an AST.
To solve the above issues, we refine the mappings of AST nodes into mappings of statements and tokens.
Moreover, we treat each statement (including the statement itself and its tokens) as an analysis unit.

\begin{figure}
  \centering
  \includegraphics[width=0.48\textwidth]{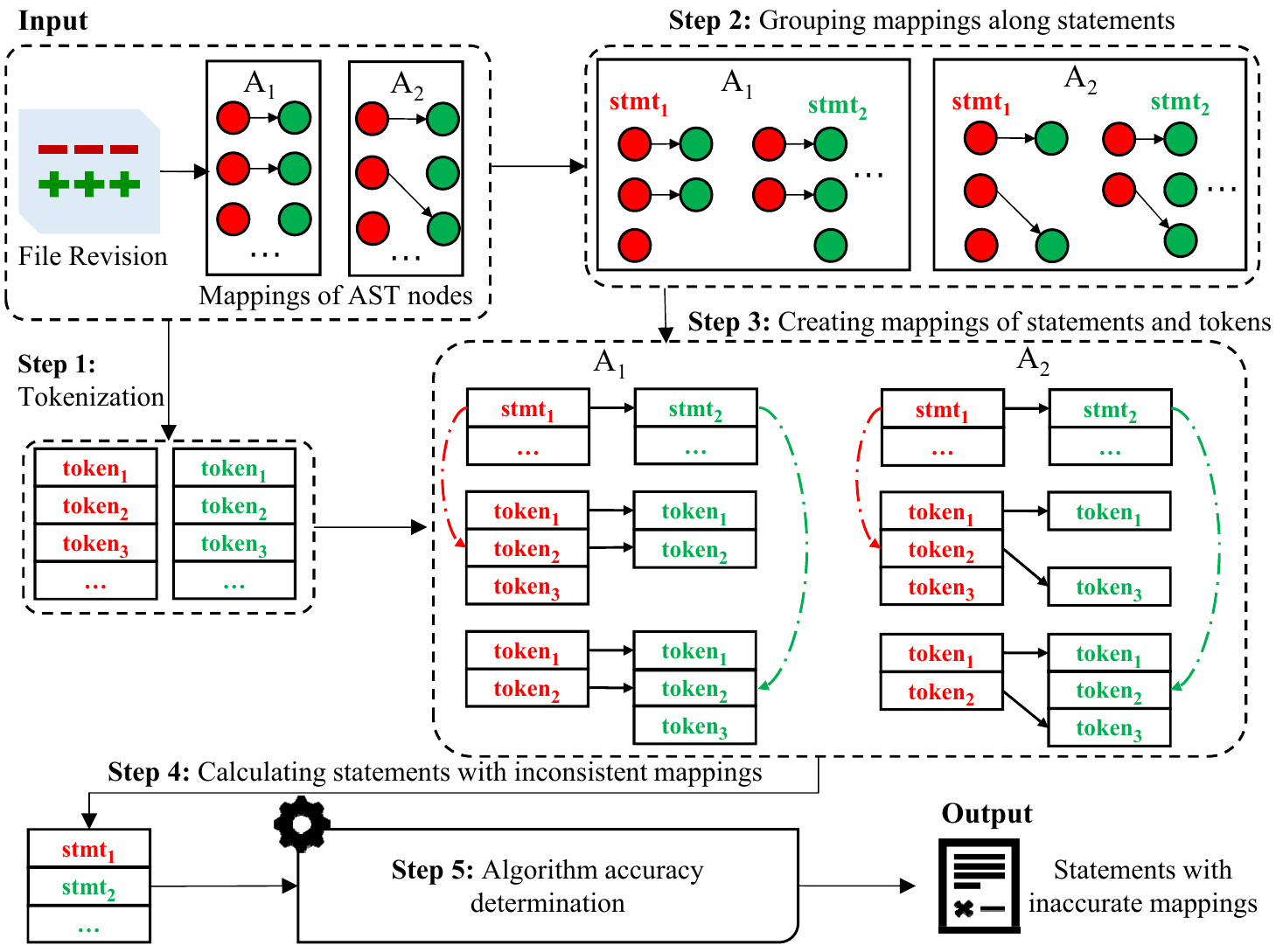}\\
  \caption{Overview of our approach. The red color and green color denote that the program element is from the file before and after the revision, respectively.}\label{fig:overview}
\end{figure}

Fig.~\ref{fig:overview} presents the overview of our approach.
Given the generated mappings by two algorithms ($A_1$ and $A_2$) for a file revision, we take five steps to calculate the statements with inaccurate mappings for each algorithm.
In Sections~\ref{step1} to~\ref{approach}, we describe the five steps, respectively.
In Section~\ref{sec:measure}, we elaborate how to compare the similarity of mapped statements and tokens by different algorithms.

\subsection{Tokenization}\label{step1}

In this step, we tokenize the file before and after a revision.
Instead of using punctuation and spaces, we use the parsed AST of the source file to tokenize the file.
For example, a string literal containing punctuation and spaces is considered as a single token.

For a Java file, we first use the Eclipse JDT parser to generate a standard JDT AST.
Then, we extract the tokens from value of each AST node.
Notice that we ignore AST nodes representing comments and Javadocs.
Because comments and Javadocs are typically not treated as code~\cite{huang2018cldiff}.
As a result, we retrieve two token lists for a file before and after a revision, respectively.
The token lists are not impacted by the used AST by each algorithm

\subsection{Grouping Mappings along Statements}\label{step2}
In this step, we separately group mappings of AST nodes along statements for $A_1$ and $A_2$.
In an AST, a statement (e.g., a method declaration) can have descendant statements.
For each statement in the source and target ASTs that are being analyzed by an algorithm, we first group the nodes that belong to the statement but do not belong to its descendant statements.
For instance, our framework groups $n_1$, $n_2$, and $n_3$ in Fig.~\ref{fig:mapping-example}(b) for the type declaration \texttt{PublishedAddressPolicy}.
Then, we find the generated mappings by the algorithm for the grouped nodes.
These mappings are grouped for the statement.

\subsection{Creating Mappings of Statements and Tokens}\label{step3}

In this step, we separately calculate mappings of statements and tokens for $A_1$ and $A_2$.
Among the grouped mappings for each statement, we can directly find the mapping for the statement.
For instance, in Fig.~\ref{fig:mapping-example}(b), $n_1$ represents the type declaration \texttt{PublishedAddressPolicy} and the mapping of $n_1$ is considered as the mapping of the declaration.

For a token, we refer to an AST node whose corresponding program element contains the token as a \emph{relevant node} of the token.
Among the relevant nodes of a token, we refer to the node of the lowest level as the \emph{directly relevant node} of the token.
In return, we refer to the token as a \emph{directly relevant token} of the node.
We take the first \texttt{HashMap} of line 1 in Fig.~\ref{fig:mapping-example}(a) as an example.
In the used AST by GT shown in Fig.~\ref{fig:mapping-example}(b), $n_1$, $n_4$, $n_6$, $n_7$ and $n_8$ are relevant nodes of the token.
And the directly relevant node for the token is $n_8$.
In the used AST by IJM shown in Fig.~\ref{fig:mappings-ijm}(a), $n_1$, $n_3$ and $n_5$ are relevant nodes of the token.
And the directly relevant node for the token is $n_5$.

An AST node can have several directly relevant tokens.
For instance, the node $n_5$ in Fig.~\ref{fig:mappings-ijm}(a) has three directly relevant tokens including a \texttt{HashMap} and two \texttt{Integer} tokens.
Such tokens compose the value of $n_5$.
Another example is $n_1$ in Fig.~\ref{fig:mappings-ijm}, which has two directly relevant tokens including the \texttt{class} and \texttt{PublishedAddressPolicy} tokens in Fig.~\ref{fig:mapping-example}(a).
The \texttt{PublishedAddressPolicy} token is the value of the node, while the \texttt{class} token does not belong to its value.

We observe that a token's mapping is determined by the mapping of its directly relevant node.
We also take the first \texttt{HashMap} of line 1 in Fig.~\ref{fig:mapping-example}(a) as an example.
In Fig.~\ref{fig:mapping-example}, GT maps $n_8$ to $n_{59}$---indicating that the token is mapped to the \texttt{HashMap} of line 3 in Fig.~\ref{fig:mapping-example}(a).
And in Fig.~\ref{fig:mappings-ijm}, IJM maps $n_5$ to $n_{13}$---indicating that the token can only be mapped to a directly relevant token of $n_{13}$.

For each node of the source and target ASTs, we calculate all the directly relevant tokens and list the tokens according to their character positions.
Then, for each pair of mapped nodes, we map tokens from the lists of directly relevant tokens of the nodes.
If both lists have only one token, we directly map two tokens.
Otherwise, we separately map the tokens composing the value of the nodes and other tokens, since tokens composing the value of a node can only be mapped to those composing the value of another node.
For the tokens composing the values of the two nodes, we first sequentially map identical tokens between the two lists and then map the tokens (including the first and last tokens) that are surrounded by already mapped pairs of tokens.
Mapping program elements surrounded by already mapped pairs is a commonly used heuristic by program element mapping algorithms~\cite{kim2006program}.
After that, our framework applies the same method to map the tokens that do not belong to the value of the two nodes.

For instance, in Fig.~\ref{fig:mappings-ijm}, $n_5$ and $n_{13}$ are mapped.
The directly relevant tokens of $n_5$ include \texttt{HashMap}, \texttt{Integer} and \texttt{Integer}.
And the directly relevant tokens of $n_{13}$ include \texttt{Map}, \texttt{Integer} and \texttt{Integer}.
All of the tokens belong to the values of the two nodes.
We first sequentially map the two \texttt{Integer} tokens between the two lists.
\texttt{HashMap} and \texttt{Map} are surrounded by already mapped tokens and they are further mapped.

As a result, we calculate mappings of tokens using the generated mappings of the AST nodes by $A_1$ and $A_2$.
We further group mappings of tokens along each statement based on the grouped mappings of AST nodes for the statement.

\subsection{Calculating Statements with Inconsistent Mappings}\label{step4}

In this step, we calculate statements with inconsistent mappings for comparing $A_1$ and $A_2$.
For each statement in the file before and after the revision, we first calculate whether the two algorithms inconsistently map the statement.
Then, for each token in the statement, we calculate whether the two algorithms inconsistently map the token.
Finally, we output statements with inconsistent mappings and for each statement, we group the inconsistent mappings of the statement and its tokens by comparing the two algorithms.

\subsection{Algorithm Accuracy Determination}\label{approach}

In this step, we compare the similarity of mapped statements and tokens across each pair of different algorithms.
By performing the comparison, we determine the accuracy of each algorithm in the mapping of a statement or a token.
In this section, we introduce how to determine the accuracy of each algorithm by performing the comparison.
In Section~\ref{sec:measure}, we elaborate how to compare the similarity of mapped statements and tokens by different algorithms.

Let us denote the similarity between two program elements $e$ and $\hat{e}$ as $Sim(e, \hat{e})$.
The elements can be statements or tokens.
Suppose that two algorithms ($A_0$ and $A_1$) map a program element $e_0$ to two different elements $e_1$ and $e_2$, respectively.
We notice that the two algorithms also inconsistently map $e_1$ and $e_2$.
Suppose that $A_0$ maps an element $e_3$ to $e_2$, and $A_1$ maps an element $e_4$ to $e_1$.
In other words, $A_0$ generates two mappings $\{\langle e_0, e_1\rangle, \langle e_3, e_2\rangle\}$.
And $A_1$ generates two mappings $\{\langle e_0, e_2\rangle, \langle e_4, e_1\rangle\}$.

Suppose that $Sim(e_0, e_1)$ is larger than $Sim(e_0, e_2)$, we determine that $A_0$ is more accurate than $A_1$ in mapping $e_0$.
However, it is not enough to determine that the mapping of $e_0$ to $e_1$ is better than the generated mappings by $A_1$.
We must also check the condition: $Sim(e_0, e_1) > Sim(e_4, e_1)$, i.e., $A_0$ is also more accurate than $A_1$ in mapping $e_1$.
If this condition is also satisfied, we determine that $A_1$ inaccurately maps $e_0$ and $e_1$.
Similarly, we can also determine if the two generated mappings by $A_0$ are inaccurate.

Given a pair of algorithms and a file revision, we calculate the statements with inconsistent mappings.
For each statement with inconsistent mappings, we perform the above comparison for the mappings of the statement and its tokens as generated by the two algorithms that are being compared.
Consequently, we calculate statements with inaccurate mappings for each algorithm.

For each studied algorithm, we separately compare it with the other two studied algorithms.
Finally, we calculate a union set of statements with inaccurate mappings for any of the algorithms that are being compared.

\subsection{Similarity Comparison}\label{sec:measure}

Our aim is to automatically compare the similarity of mapped statements and tokens by different algorithms.
To realize this aim, we perform a manual analysis of statements with inconsistent mappings.
Also, we verify if statements with inconsistent mappings can expose the inaccurate mappings as generated by the algorithms.


In this study, we analyze ten open-source Java projects.
Table~\ref{tab:projects} presents statistics of the ten studied projects.
These projects were analyzed by prior studies~\cite{huang2018cldiff, frick2018generating}.
We collect the commits of the projects from the creation date of the projects to January 2019.

\begin{table}
  \centering
  \caption{Statistics of the studied projects.}
    \begin{tabular}{|l|r|r|}
    \hline
    \textbf{Projects} & \multicolumn{1}{l|}{\textbf{\#Commit}} & \multicolumn{1}{l|}{\textbf{\#File Revision}} \\
    \hline
    ActiveMQ &                8,059  &                24,813  \\
    Commons IO &                1,067  &                  2,727  \\
    Commons Lang &                3,031  &                  6,917  \\
    Commons Math &                4,257  &                18,133  \\
    Junit4 &                1,241  &                  3,802  \\
    Hibernate ORM &              10,170  &                51,711  \\
    Hibernate Search &                5,369  &                27,002  \\
    Spring Framework &              14,754  &                68,413  \\
    Spring Roo &                4,274  &                19,894  \\
    Netty &              11,135  &                39,753  \\
    \hline
    \textbf{Total} &              63,357  &              263,165  \\
    \hline
    \end{tabular}%
  \label{tab:projects}%
\end{table}%

We can compare three pairs of algorithms, namely GT vs. MTD, GT vs. IJM and MTD vs. IJM.
For each pair of algorithms, we sample 50 file revisions for which the two algorithms inconsistently map program elements from the studied projects.
For each file revision, we analyze all the statements before and after the revision that involve the inconsistent mappings.
We analyze 178, 191 and 206 statements for comparing the three pairs of algorithms, respectively.
In total, we analyze 575 statements.
For each statement, we determine the accuracy of mappings as generated by each of compared algorithms using Frick et al.'s criteria~\cite{frick2018generating}.
In total, we make 1,150 ($575 \times 2$) determinations.

The first and second author of this paper separately analyzed the statements with inconsistent mappings.
Both authors have at least three years of programming experience in Java.
We calculate Fleiss' Kappa~\cite{fleiss1971measuring} to estimate the agreement of the two annotators' determination results.
The Kappa value is 0.81, which  indicates an excellent agreement.
Finally, the two annotators compared their determination results to uncover disagreements.
For each statement with a disagreement, the annotators further discussed the accuracy of the generated mappings by the two compared algorithms.

We analyze all the statements with inaccurate mappings to design similarity measures to distinguish accurate and inaccurate mappings of statements and tokens.
We categorize the statements with inaccurate mappings for the algorithms along each similarity measure.
For each statement with inaccurate mappings, we consider if we have designed a measure that can identify the inaccurate mappings.
If we have designed such a measure, we categorize the statement to the category of the measure.
Otherwise, we try to design a new similarity measure to distinguish the accurate and inaccurate mappings.
By doing so, we design hierarchical similarity measures for two statements and two tokens.


From our manual analysis, we have the following findings.

\begin{itemize}
\item Among the 178 statements for comparing GT and MTD, 71 and 129 statements involve inaccurate mappings as generated by GT and MTD, respectively.
Among the 191 statements for comparing GT and IJM, 114 and 91 statements involve inaccurate mappings as generated by GT and IJM, respectively.
Among the 206 statements for comparing MTD and IJM, 113 and 117 statements involve inaccurate mappings as generated by MTD and IJM, respectively.

\item For each statement with inconsistent mappings for comparing two algorithms, at least one algorithm is determined to generate inaccurate mappings. Hence, statements with inconsistent mappings can expose the inaccurate mappings as generated by the algorithms.

\item There exist cases where two algorithms consistently produce an inaccurate mapping of a statement or a token.
\end{itemize}

\begin{table*}
  \centering
   \caption{The similarity measures for statements and tokens.}
\begin{tabular}{|c|c|>{\raggedright}p{0.5\textwidth}|r|r|r|}
\hline
\textbf{Element} & \textbf{Measure} & \textbf{Measure Description} & \textbf{GT} & \textbf{MTD} & \textbf{IJM}\tabularnewline
\hline
\hline
\multirow{2}{*}{Stmt.} & NIT & Number of mapped identical tokens between a pair of mapped statements. & 22 & 72 & 17\tabularnewline
\cline{2-6} \cline{3-6} \cline{4-6} \cline{5-6} \cline{6-6}
 & PM & Whether the parent nodes of a pair of mapped statements are mapped. & 9 & 18 & 7\tabularnewline
\hline
\multirow{4}{*}{Token} & TYPE & Whether mapped tokens have the same type. & 29 & 24 & 0\tabularnewline
\cline{2-6} \cline{3-6} \cline{4-6} \cline{5-6} \cline{6-6}
 & STMT & Whether mapped tokens belong to a pair of mapped statements. & 71 & 84 & 146\tabularnewline
\cline{2-6} \cline{3-6} \cline{4-6} \cline{5-6} \cline{6-6}
 & VAL & Whether mapped tokens have the same value. & 16 & 0 & 4\tabularnewline
\cline{2-6} \cline{3-6} \cline{4-6} \cline{5-6} \cline{6-6}
 & LLCS & Length of the longest common subsequence calculated using mapped tokens between mapped statements. & 10 & 10 & 4\tabularnewline
\hline
\multicolumn{3}{|c|}{Other} & 28 & 34 & 30\tabularnewline
\hline
\end{tabular}
\label{tab:measure}
\end{table*}

In total, we find 185 ($71+114$), 242 ($129+113$), 208 ($91 + 117$) statements with inaccurate mappings for GT, MTD and IJM, respectively.
Based on our observation, we design two similarity measures for comparing two statements and four similarity measures for comparing two tokens.
Table~\ref{tab:measure} presents the six measures.
Notice that we define two tokens with the same value as \emph{identical tokens}.
We use the six similarity measures collectively to compare the generated mappings by different algorithms.
In the remaining part of this section, we first introduce the six measures and our categorization results for the statements with inaccurate mappings.
Then we describe the usage of the proposed measures.

The six similarity measures are elaborated below.

\begin{figure}
  \centering
  \includegraphics[width=0.48\textwidth]{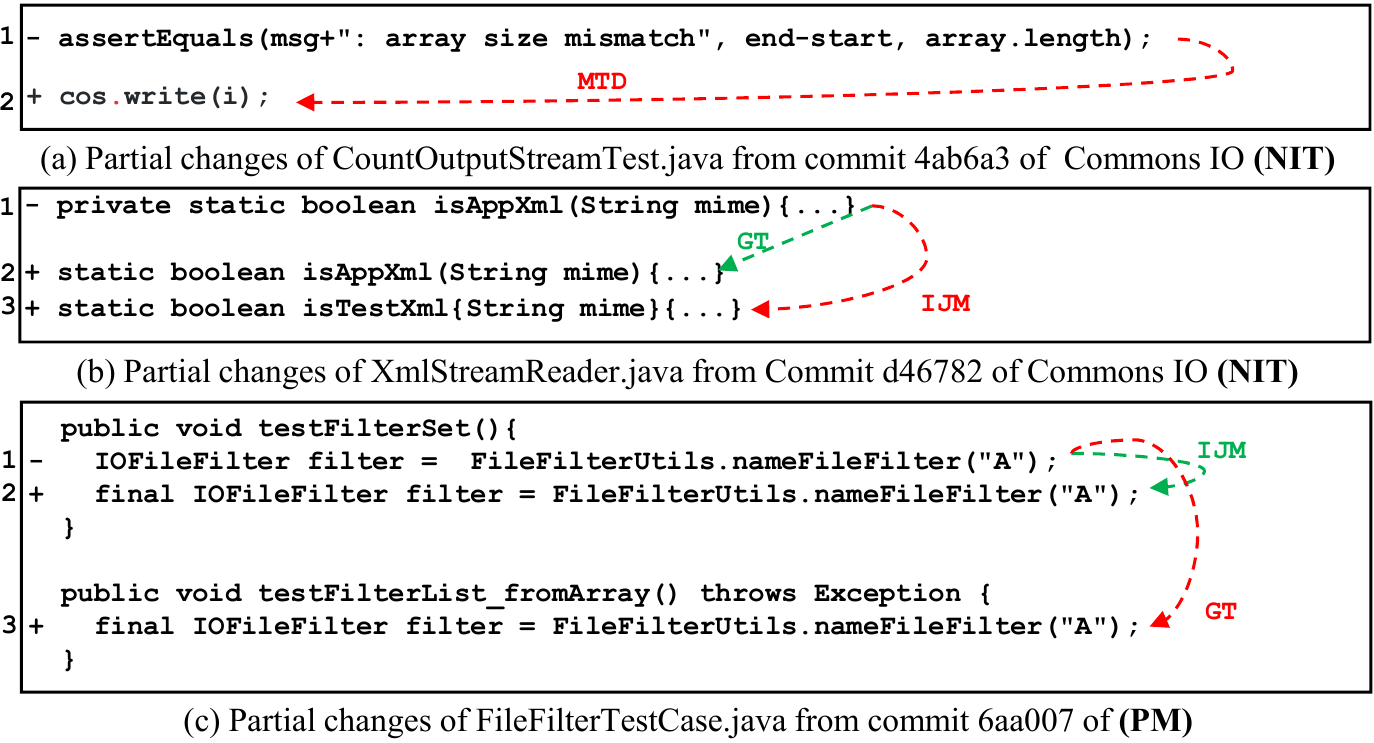}\\
  \caption{Illustrative examples of inaccurately mapped statements that can be identified using our measures.}\label{fig:measure1}
\end{figure}

\underline{\emph{NIT}} is defined as the number of mapped identical tokens between a pair of mapped statements.
Two mapped statements with a larger NIT are more similar and an NIT of 0 indicates that the two statements are highly dissimilar.
If two statements are mapped with an NIT of 0, we determine the mapping as inaccurate.
Otherwise, for a pair of mapped statements, we check if the mapping of one of the statements to another statement can achieve a larger NIT.
For instance, in Fig.~\ref{fig:measure1}(a), MTD inaccurately maps the statements at lines 1 and 2, and the NIT is 0.
In Fig.~\ref{fig:measure1}(b), GT maps the statements at lines 1 and 2 with an NIT of five.
IJM maps the statements at lines 1 and 3 with an NIT of four.
We determine that GT is more accurate than IJM in mapping the statement at line 1.

\underline{\emph{PM}} characterizes whether the parent nodes of a pair of mapped statements are also mapped.
We observe that statements with mapped parent nodes are more likely to be mapped.
For a pair of mapped statements, we check if mapping one of the statements to another statement with mapped parent nodes can achieve the same NIT.
For instance, in Fig.~\ref{fig:measure1}(c), IJM maps the statements at lines 1 and 2 with mapped parent nodes, while GT maps the statements at lines 1 and 3 with parent nodes not mapped.
We determine that IJM is more accurate than GT in mapping the statement at line 1.
Furthermore, we notice a special type of statements, i.e., blocks.
A block is a group of statements between balanced braces (i.e., ``\{'' and ``\}'').
We observe that a block should be mapped along with its parent nodes, e.g., the ``\{'' following the method \texttt{testFilterSet} in Fig.~\ref{fig:measure1}(c) should be mapped along with the method declaration.
Thus, mapped blocks with unmapped parent nodes are determined to be inaccurate.

\begin{figure}
  \centering
  \includegraphics[width=0.48\textwidth]{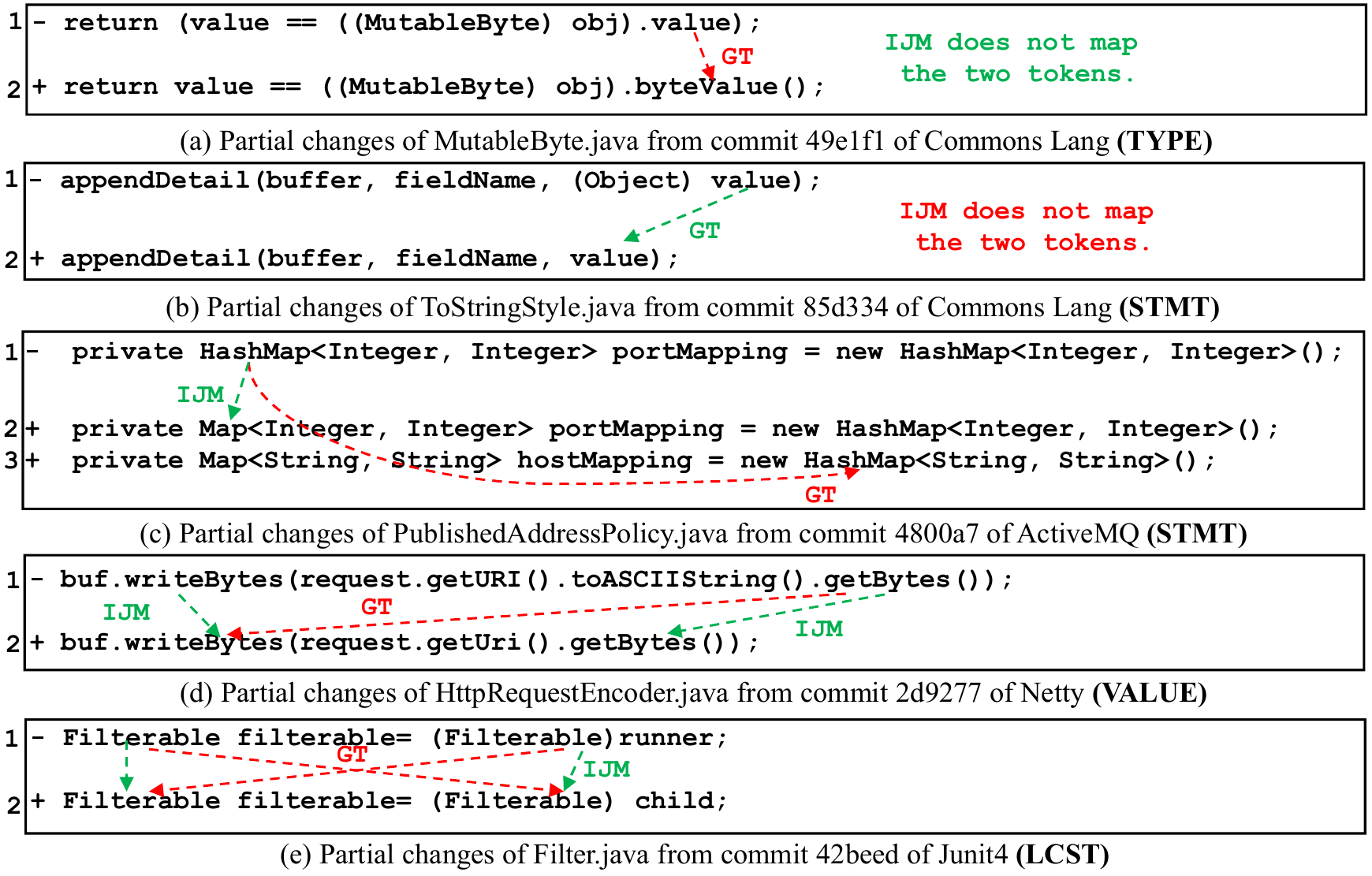}\\
  \caption{Illustrative examples of inaccurately mapped tokens that can be identified using our measures.}\label{fig:measure2}
\end{figure}

\underline{\emph{TYPE}} characterizes whether mapped tokens have the same type.
For a token whose directly relevant node is not a name node, we define the type of the token as the label of its directly relevant node.
For tokens whose directly relevant node is a name node, we define four types: variable name, type name, method name and declaration name.
Following Frick et al.~\cite{frick2018generating}, we consider the mapping of tokens with different types as inaccurate.
For instance, in Fig.~\ref{fig:measure2}(a), GT maps a variable name \texttt{value} to a method name \texttt{bytevalue}, we determine that such a mapping is inaccurate.


\underline{\emph{STMT}} characterizes whether mapped tokens belong to a pair of mapped statements.
Two tokens from mapped statements are more likely to be mapped.
We observe that mapping tokens from mapped statements is better than (1) not mapping the tokens and (2) mapping tokens from unmapped statements.
For instance, in Fig.~\ref{fig:measure2}(b), the two \texttt{value} tokens are both variable names and they belong to a pair of mapped statements.
GT maps the two tokens, while IJM does not map them.
We determine that GT is more accurate than IJM in mapping the two tokens.
As another example, we find that both GT and IJM map the statement at line 1 to the statement at line 2 in Fig.~\ref{fig:measure2}(c).
GT maps the two \texttt{HashMap} tokens in unmapped statements, while IJM maps the \texttt{HashMap} to the \texttt{Map} from the mapped statements.
Using the STMT measure, we determine that IJM is more accurate in mapping the \texttt{HashMap} at line 1 than GT.

\underline{\emph{VAL}} characterizes whether mapped tokens have the same string value.
Two identical tokens in a pair of mapped statements are more likely to be mapped.
We consider that mapping such two tokens is better than mapping one of the tokens to another token with different values between the two statements.
For instance, in Fig.~\ref{fig:measure2}(d), GT maps \texttt{getBytes} to \texttt{writeBytes}, while IJM maps the two \texttt{getBytes} tokens.
Using the VAL measure, we determine that IJM is more accurate in mapping the \texttt{getBytes} tokens than GT.

\underline{\emph{LLCS}} is defined as the length of the longest common subsequence (LCS)~\cite{hirschberg1977algorithms} that is calculated using the mapped tokens between mapped statements.
We observe that the order of tokens is infrequently changed in a statement.
We use LLCS to quantify the number of tokens that are sequentially mapped between the mapped statements.
For instance, in Fig.~\ref{fig:measure2}(e), GT changes the orders of the two \texttt{Filterable} tokens in the statement at line 1.
As a result, at most three tokens are sequentially mapped between the statements, i.e., \texttt{Filterable}, \texttt{=} and \texttt{runner}.
The LLCS for the mapped tokens is calculated as three.
IJM sequentially maps the five tokens between the two statements.
The LLCS for the mapped tokens is calculated as five.
Using the LLCS measure, we determine that IJM is more accurate in mapping the \texttt{Filterable} tokens than GT.

Table~\ref{tab:measure} shows the number of statements that are categorized along the measures.
The measures can identify 157 (85\%), 208 (86\%) and 178 (86\%) of the statements with inaccurate mappings for GT, MTD and IJM, respectively.
The other statements with inaccurate mappings are categorized into the \textbf{Other} category.
For these statements, we find that determining the accuracy of the mappings of statements and tokens requires more comprehension of the changes.

\noindent\textbf{Usage of the similarity measures.} We compare the generated mappings by two algorithms using the following steps.

\noindent\underline{\emph{Step 1.}}
If an algorithm maps two non-block statements with an NIT of 0, we determine the mapping as inaccurate.
If an algorithm maps two blocks with unmapped parent nodes, we determine the mapping as inaccurate.
If an algorithm maps two tokens with different types, we determine the mapping as inaccurate.

\noindent\underline{\emph{Step 2.}}
When the two algorithms map a statement to different statements, mapping statements with a larger NIT is considered to be more accurate.
If the two pairs of statements have the same NIT, mapping statements with mapped parent nodes is considered to be more accurate than mapping statements with unmapped parent nodes.

\noindent\underline{\emph{Step 3.}}
When the two algorithms consistently map a statement, we assume that the two algorithms accurately map the statement.
From the statement, we retrieve all the tokens that are inconsistently mapped by the two algorithms.
For a token that is inconsistently mapped by the two algorithms, we first use the STMT measure to compare the generated mappings for the token by the two algorithms.
If both algorithms map tokens from the mapped statements, mapping identical tokens is considered to be more accurate than mapping tokens with different values.
If both algorithms map identical tokens from the mapped statements, the mapped tokens with larger LLCS are considered to be more accurate.

\section{Evaluation}\label{results}

We evaluate our approach by answering three research questions.
In this section, we present the three research questions and our answer to each question.

\subsection{(RQ1) How effective is our approach in detecting statements with inaccurate mappings for the studied algorithms?}

\noindent\textbf{Motivation.}
By answering this research question, we investigate if our approach can effectively find the statements with inaccurate mappings as generated by the studied algorithms.

\noindent\textbf{Method.}
Our approach may be over-fit on the used dataset in our manual analysis.
Thus, we conduct an experiment with 12 external experts.
The experts include PhD students and post-doctors majoring in software engineering.
They have three to seven years of programming experience in Java.
Seven experts have prior experience working in industry.
For each project, we randomly select 20 statements from all the file revisions.
For each selected statement, at least two studied algorithms inconsistently map the statement or its tokens.
In total, we select 200 of such statements with inconsistent mappings.
The selected statements involve various change patterns including adding, deleting, moving and updating statements and tokens.

We randomly divide the 200 statements into four groups with each group having 50 statements.
We also divide the experts into four groups with each group having three experts.
We invite the four groups of experts to analyze the four groups of statements, respectively.
For each statement, we provide the mappings of the statement and its tokens as generated by each of the studied algorithms.
Notice that we do not provide the algorithm that generates the mappings.
We let the experts determine if the mapping of the statement or a token of the statement is inaccurate.
For each group of statements, we calculate Fleiss' Kappa~\cite{fleiss1971measuring} to estimate the agreement of the three experts' determination results.

For each studied algorithm, we have three determination results on the accuracy of algorithm in mapping each statement and its tokens.
For each statement and the generated mappings of the statement and its tokens by an algorithm, we label the mappings as inaccurate if at least two experts determine that inaccurate mappings exist.

Then, we run our approach to determine statements with inaccurate mappings for GT, MTD and IJM from the 200 statements.
Finally, we compare the determination results of our approach with the experts' determination results.
We define a true positive as a statement with inaccurate mappings for an algorithm that is determined as such by both our approach and experts.
We define a false positive as a statement with inaccurate mappings for an algorithm that is determined as such by our approach but not determined as such by experts.
We define a false negative as a statement with inaccurate mappings for an algorithm that is determined as such by experts but not determined as such by our approach.
Let us denote the number of true positives, false positives and false negatives as TP, FP and FN.
We calculate the precision of our approach as $\frac{TP}{TP+FP}$.
And we calculate the recall of our approach as $\frac{TP}{TP+FN}$.

\noindent\textbf{Results.}
For the four groups of statements, the Kappa values for the experts' determination results are 0.81, 0.82, 0.84 and 0.78, respectively.
Thus, the experts' determination results have an excellent agreement.

\begin{table}
  \centering
  \caption{TP, FP, FN, precision and recall of our approach.}
    \begin{tabular}{|l|r|r|r|r|r|}
    \hline
    \textbf{Alg.} & \multicolumn{1}{l|}{\textbf{TP}} & \multicolumn{1}{l|}{\textbf{FP}} & \multicolumn{1}{l|}{\textbf{FN}} & \multicolumn{1}{l|}{\textbf{Precision}} & \multicolumn{1}{l|}{\textbf{Recall}} \\
    \hline
    \hline
    GT    & 56    & 1     & 27    & 0.98  & 0.67 \\
    MTD   & 90    & 0     & 30    & 1.00  & 0.75 \\
    IJM   & 59    & 1     & 32    & 0.98  & 0.65 \\
    \hline
    \end{tabular}%
  \label{tab:rq1}%
\end{table}%

\begin{figure*}
  \centering
  \includegraphics[width=0.8\textwidth]{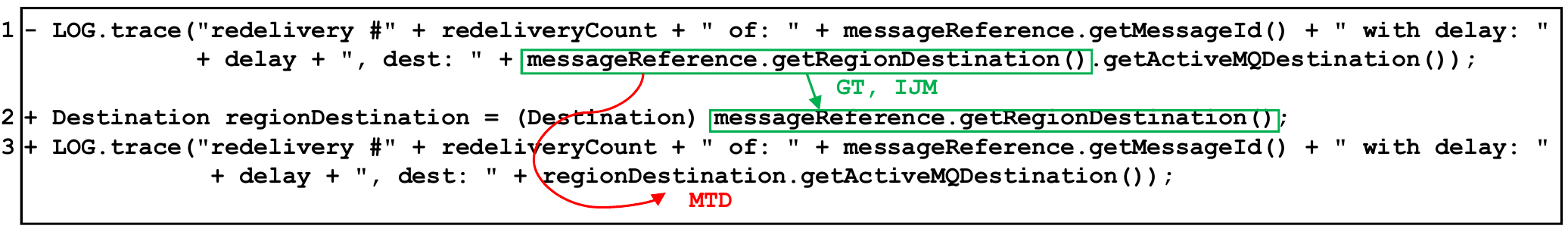}\\
  \caption{GT and IJM generates accurate mappings but our approach determines the mapping of a token as inaccurate.}\label{fig:err-example}
\end{figure*}

Table~\ref{tab:rq1} presents the TP, FP, FN, precision and recall of our approach in determining statements with inaccurate mappings for the studied algorithms.
As shown in the table, our approach achieves a precision of 0.98--1.00 and a recall of 0.65--0.75.
Almost all of the statements with inaccurate mappings as determined by our approach are also determined as such by experts.

For the false positives and false negatives, we further asked the experts why they considered that an algorithm inaccurately maps a statement or tokens of the statement.
We analyze cases of false positives and false negatives.
For the two false positives, we find that GT and IJM generate the accurate mappings for a statement and its tokens but our approach determines the mapping of a token as inaccurate.
We show this case in Fig.~\ref{fig:err-example}.
As shown in Fig.~\ref{fig:err-example}, the code involves a refactoring that extracts the method invocation \texttt{messageReference.getRegionDestination} in the statement at line 1 as a new variable.
GT and IJM accurately map the invocation to the statement at line 2, while MTD maps \texttt{messageReference} in the statement at line 1 to the \texttt{regionDestination} in the statement at line 3.
In this case, mapping tokens from unmapped statements is better than mapping tokens from mapped statements.
However, when comparing the generated mappings by GT, IJM and MTD, our approach considers that MTD generates a better mapping than GT and IJM.
This case indicates that our approach can be further improved by considering refactoring changes.

For 11 cases of false negatives, we find that our similarity measures can distinguish the accurate and inaccurate mappings of statements or tokens.
However, all three algorithms generate inaccurate mappings.
Hence, the inaccurate mappings cannot be detected by comparing the similarity measures of the mapped statements and tokens between different algorithms.
We observe 38 cases where an algorithm maps two tokens from unmapped statements and another algorithm separately maps the two tokens to empty elements.
We observe 33 cases where an algorithm maps two statements and another algorithm separately maps the two statements to empty elements.
We further observe 7 cases where two algorithms map a statement or token to different statements or tokens but our similarity measures cannot distinguish accurate and inaccurate mappings.
In these 78 cases, determining the inaccurate algorithm requires more syntactic information to determine if mapping two tokens or two statements helps understand the changes.

\noindent\textbf{Summary.}
Our approach achieves a precision of 0.98--1.00 and a recall of 0.65--0.75 in determining the statements with inaccurate mappings for the studied algorithms.
Any statements with inaccurate mappings that we detect are highly likely to be correct, although there may be additional inaccurate mappings that we cannot detect.
Our approach can be used to estimate the lower bound on the effectiveness of AST mapping algorithms.

\subsection{(RQ2) How effective is our approach when comparing an algorithm with multiple algorithms than when comparing it with another algorithm?}

\noindent\textbf{Motivation.}
As described in Section~\ref{approach}, we separately compare an algorithm with the other two algorithms.
Then, we calculate a union set of statements with inaccurate mappings for the algorithm.
We investigate if comparing an algorithm with the other two algorithms is more effective in detecting statements with inaccurate mappings than comparing it with another algorithm.

\noindent\textbf{Method.}
We have three pairs of studied algorithms, namely GT vs. MTD, GT vs. IJM and MTD vs. IJM.
For the 200 analyzed statements in RQ1, we use our approach to compare the generated mappings for statements and tokens by each pair of algorithms.
For each pair of algorithms, we calculate a set of statements with inaccurate mappings.
In such a case, we compare an algorithm with another algorithm.
Then, we calculate the precision and recall of our approach in detecting statements with inaccurate mappings for the two algorithms.
The precision and recall of our approach that compares an algorithm with the other two algorithms are shown in Table~\ref{tab:rq1}.
Finally, we compare the results shown in the table with the precision and recall of our approach that compares an algorithm with another algorithm.

\begin{table}
  \centering
  \caption{TP, FP, FN, precision and recall of our approach when comparing an algorithm with another algorithm.}
    \begin{tabular}{|c|l|r|r|r|r|r|}
    \hline
    \multicolumn{1}{|l|}{\textbf{Comparison}} & \textbf{Alg.} & \multicolumn{1}{l|}{\textbf{TP}} & \multicolumn{1}{l|}{\textbf{FP}} & \multicolumn{1}{l|}{\textbf{FN}} & \multicolumn{1}{l|}{\textbf{Precision}} & \multicolumn{1}{l|}{\textbf{Recall}} \\
    \hline
    \hline
    \multirow{2}[0]{*}{GT vs. MTD} & GT    & 44    & 1     & 39    & 0.98  & 0.53 \\
          & MTD   & 83    & 0     & 37    & 1.00  & 0.69 \\
    \hline
    \multirow{2}[0]{*}{GT vs. IJM} & GT    & 47    & 0     & 36    & 1.00  & 0.57 \\
          & IJM   & 51    & 0     & 40    & 1.00  & 0.56 \\
    \hline
    \multirow{2}[0]{*}{MTD vs. IJM} & MTD   & 74    & 0     & 46    & 1.00  & 0.62 \\
          & IJM   & 49    & 1     & 42    & 0.98  & 0.54 \\
    \hline
    \end{tabular}%
  \label{tab:rq2}%
\end{table}%

\begin{table*}
  \centering
  \caption{Number of statements and file revisions for which the studied algorithms are determined to generate inaccurate mappings.}
    \begin{tabular}{|l|r|r|r|r|r|r|}
    \hline
    \multicolumn{1}{|c|}{\multirow{2}[0]{*}{\textbf{Projects}}} & \multicolumn{3}{c|}{\textbf{Statements }} & \multicolumn{3}{c|}{\textbf{File Revisions}} \\
    \cline{2-7}
          & \multicolumn{1}{l|}{\textbf{GT}} & \multicolumn{1}{l|}{\textbf{MTD}} & \multicolumn{1}{l|}{\textbf{IJM}} & \multicolumn{1}{l|}{\textbf{GT}} & \multicolumn{1}{l|}{\textbf{MTD}} & \multicolumn{1}{l|}{\textbf{IJM}} \\
    \hline
    \hline
    ActiveMQ &          53,083  &           191,566  &          39,669  &          5,817  &          8,523  &          5,786  \\
    Commons IO &           7,932  &            16,978  &           5,883  &            546  &            713  &            656  \\
    Commons Lang &          23,306  &            53,533  &          21,567  &          1,501  &          1,823  &          1,641  \\
    Commons Math &          43,450  &           101,194  &          34,440  &          3,703  &          4,588  &          3,881  \\
    Junit4 &           6,997  &            13,800  &           5,750  &            924  &          1,083  &            984  \\
    Hibernate ORM &        127,770  &           356,412  &          93,146  &        13,026  &        16,414  &        13,069  \\
    Hibernate Search &          43,942  &           112,419  &          38,919  &          6,012  &          7,326  &          6,964  \\
    Spring Framework &        164,545  &           440,480  &        145,374  &        17,120  &        20,484  &        20,215  \\
    Spring Roo &          54,245  &           172,454  &          38,397  &          4,573  &          5,764  &          4,963  \\
    Netty &        130,249  &           374,883  &          91,795  &        11,584  &        14,191  &        11,451  \\
    \hline
    \end{tabular}%
  \label{tab:rq3}%
\end{table*}%

\noindent\textbf{Results.}
Table~\ref{tab:rq2} presents TP, FP, FN, precision and recall of our approach when comparing an algorithm with another algorithm.
By comparing the results show in Tables~\ref{tab:rq1} and~\ref{tab:rq2}, we find that our approach achieves a better recall with a difference of 9\%--23\% when comparing an algorithm with two algorithms than when comparing it with another algorithm.
On the other hand, the precision of our approach is not impacted.

As described in Section~\ref{sec:measure}, two algorithms may generate the same mapping that is inaccurate.
Such an inaccurate mapping cannot be detected by comparing the two algorithms.
If another algorithm generates the accurate mapping, comparing the third algorithm with the former two algorithms may reveal the inaccurate mapping.
Thus, comparing an algorithm with multiple algorithms can detect more inaccurate mappings.

\noindent\textbf{Summary.}
Our approach can detect 9\%--23\% more statements with inaccurate mappings when comparing an algorithm with the other two algorithms than when comparing it with another algorithm.


\subsection{(RQ3) Do state-of-the-art AST mapping algorithms generate many inaccurate mappings?}

\noindent\textbf{Motivation.}
We show that our approach achieves a nearly perfect precision in finding statements with inaccurate mappings for the studied algorithms.
Hence, we leverage our approach to investigate whether the studied algorithms generate many inaccurate mappings.

\noindent\textbf{Method.}
We leverage GT, MTD and IJM to calculate the mappings of the AST nodes for all the file revisions of the ten studied projects.
For each file revision, we use our approach to detect the statements with inaccurate mappings for each studied algorithm.
For each project, we count the detected statements with inaccurate mappings for each algorithm.
We also count the file revisions for which the studied algorithms are determined to generate inaccurate mappings.

\noindent\textbf{Results.}
Table~\ref{tab:rq3} presents the number of statements with inaccurate mappings as detected by our approach.
We also show the number of file revisions for which the studied algorithms are determined to generate inaccurate mappings.
As shown in the table, the three studied algorithms may generate a considerable number of inaccurate mappings.
For each project, we further calculate the ratio of file revisions for which the studied algorithms are determined to generate inaccurate mappings.
We find that GT, MTD and IJM are determined to generate inaccurate mappings for 20\%--29\%, 25\%--36\% and 21\%--30\% of the file revisions, respectively.

 \noindent\textbf{Summary.}
GT, MTD and IJM are determined to generate inaccurate mappings for a considerable number of file revisions.
State-of-the-art AST mapping algorithms still have room for improvement.

\section{Discussion}\label{discussion}

\subsection{Threats to Validity}

The primary threats to the validity of our experiments are twofold.
First, we compare the determination results of our approach and experts on the accuracy of generated mappings by the studied algorithms for 200 statements.
The number of analyzed statements is not very large-scale.
This is because such a manual analysis is time-consuming, with understanding mappings of each statement and each token.
On average, each expert takes 1.5 hours to analyze the allocated 50 statements.
The 200 statements are randomly taken from 10 different projects, and they are from different file revisions.
Dotzler et al. analyzed only 10 file revisions when evaluating MTD~\cite{dotzler2016move}.
Our analysis involves much more file revisions than their analysis.
Second, when we select the statements, we require that at least two studied algorithms inconsistently map the statement or its tokens.
There may exist cases where the studied algorithms consistently map a statement and its tokens, but the mapping of the statement or a token of the statement is inaccurate.
The selected statements do not consider such cases, and our approach cannot detect the inaccurate mapping in such cases.
We manually analyzed 100 statements for which the studied algorithms generate consistent mappings at both statement and token levels.
We did not observe the cases where the three algorithms produce inaccurate mappings.
Nevertheless, our code and data are made publicly available~\cite{impl}, and researchers are encouraged to investigate this possibility.

\subsection{Limitations}

From our experiments, we observe two limitations of our approach.
First, as described in our answer to RQ1, refactoring changes may impact the precision of our approach.
In refactoring changes, mapping tokens from unmapped statements may be better than mapping tokens from mapped statements.
We note that researchers proposed several refactoring detection tools, e.g.,~\cite{tsantalis2018accurate}.
Incorporating such tools into our approach may deal with this limitation.
On the other hand, there still exists a considerable number of inaccurate mappings that cannot be detected by our approach.
According to our answer to RQ2, comparing an algorithm with more algorithms may detect more inaccurate mappings as generated by the algorithm.
Moreover, researchers have proposed various heuristics to map program elements~\cite{kim2006program}.
Additional similarity measures  can be derived from these heuristics.
Such measures may further improve the recall of our approach.
Our code and data are made publicly available~\cite{impl}, and researchers are encouraged to extend our approach.

%
%
%
%

\section{Related Work}\label{related}


\subsection{AST mapping algorithms}

Many AST mapping algorithms were proposed in prior studies.
Yang proposed an AST mapping algorithm using a branch-and-bound implementation of the largest common subtree problem~\cite{yang1991identifying}.
This algorithm does not consider moved AST nodes.
Fluri et al. proposed ChangeDistiller, an AST mapping algorithm that uses a reduced AST, in which code statements are encoded as leaf nodes~\cite{fluri2007change}.
Hashimoto et al. proposed Diff/TS, an algorithm that works with raw ASTs and supports multiple languages~\cite{hashimoto2008diff}.
Nguyen et al. proposed Jsync, which leverages a classic text-based mapping algorithm to map AST nodes~\cite{nguyen2011clone}.
Recently, researchers proposed GumTree~\cite{falleri2014fine}, MTDiff~\cite{dotzler2016move} and IJM~\cite{frick2018generating}.
These algorithms are the state-of-the-art AST mapping algorithms and are analyzed in our paper.
Different from them, we focus on evaluating AST mapping algorithms rather than proposing a new AST mapping algorithm.

\subsection{Use of AST mapping algorithms}

AST mapping algorithms are widely used in several SE research areas.
ChangeDistiller has been used to identify non-essential modifications~\cite{kawrykow2011non} and automate repetitive edits~\cite{meng2013lase}.
Nguyen et al. used Jsync to track cloned code in the software evolution process~\cite{nguyen2011clone}.
Moreover, many studies used GumTree to analyze code patterns of changes such as bug-fixing changes~\cite{liu2018closer, ni2020analyzing, hanam2016discovering, campos2019discovering, islam2020bugs, koyuncu2020fixminer}, logging changes~\cite{li2018logtracker} and changes to online code examples~\cite{zhang2019analyzing}.
Also, prior work trained models based on the edit actions of changes that are calculated using GumTree~\cite{tufano2019empirical, tufano2018learning, hassan2018hirebuild, ma2017vurle, danglot2020approach}.
Such models are used to recommend changes such as patches~\cite{tufano2019empirical} and logging changes~\cite{li2018logtracker}.
Different from them, we focus on evaluating AST mapping algorithms instead of using the algorithms to analyze changes.

\section{Conclusion and Future Work}\label{conclusion}

In this paper, we propose a differential testing approach that can automatically determine the statements with inaccurate mappings for AST mapping algorithms.
Given a file revision, we first compare the generated mappings by different algorithms and extract the statements with inconsistent mappings.
Then, we use six similarity measures collectively to compare the mapped statements and tokens by different algorithms.
By doing so, we determine the statements with inaccurate mappings for each of the algorithms.

By conducting an experiment with 12 experts, we show that our approach achieves a precision of 0.98--1.00 and a recall of 0.65--0.75.
The studied algorithms are determined to generate inaccurate mappings for a considerable number (20\%--36\%) of file revisions in our studied projects.
Hence, state-of-the-art AST mapping algorithms still need improvements.
AST mapping algorithms play a foundational role in many existing studies.
It is necessary to investigate if the inaccurate mappings as generated by the algorithms impact the conclusions of existing studies.

\section*{Acknowledgement}

This research was partially supported by the National Key R\&D Program of China (No. 2019YFB1600700), Australian Research Council's Discovery Early Career Researcher Award (DECRA) funding scheme (DE200100021), ARC Discovery grant (DP200100020), Key Research and Development Program of Zhejing Province (No. 2021C01014), and the National Research Foundation, Sinapore under its Industry Alignment Fund - Prepositioning (IAF-PP) Funding Initiative. Any opinions, findings and conclusions or recommendations expressed in this material are those of the author(s) and do not reflect the views of National Research Foundation, Singapore.

\balance
\bibliographystyle{abbrv}
\bibliography{refs}

\end{document}